# Human Activity Recognition Using Tools of Convolutional Neural Networks: A State of the Art Review, Data Sets, Challenges and Future Prospects

Md. Milon Islam[1*], Sheikh Nooruddin[1], Fakhri Karray[1,2], Ghulam Muhammad[3,4]

[1]Centre for Pattern Analysis and Machine Intelligence, Department of Electrical and Computer Engineering, University of Waterloo, ON N2L 3G1, Canada
[2]Mohamed Bin Zayed University of Artificial Intelligence, Abu Dhabi, United Arab Emirates
[3]Department of Computer Engineering, College of Computer and Information Sciences, King Saud University, Riyadh, Saudi Arabia
[4]Center of Smart Robotics Research, King Saud University, Riyadh, Saudi Arabia
milonislam@uwaterloo.ca[*], sheikh.nooruddin@uwaterloo.ca, karray@uwaterloo.ca, ghulam@ksu.edu.sa

## Abstract

Human Activity Recognition (HAR) plays a significant role in the everyday life of people because of its ability to learn extensive high-level information about human activity from wearable or stationary devices. A substantial amount of research has been conducted on HAR and numerous approaches based on deep learning and machine learning have been exploited by the research community to classify human activities. The main goal of this review is to summarize recent works based on a wide range of deep neural networks architecture, namely convolutional neural networks (CNNs) for human activity recognition. The reviewed systems are clustered into four categories depending on the use of input devices like multimodal sensing devices, smartphones, radar, and vision devices. This review describes the performances, strengths, weaknesses, and the used hyperparameters of CNN architectures for each reviewed system with an overview of available public data sources. In addition, a discussion with the current challenges to CNN-based HAR systems is presented. Finally, this review is concluded with some potential future directions that would be of great assistance for the researchers who would like to contribute to this field.

## 1 Introduction

The purpose of human activity recognition is to recognize the physical tasks of a particular individual or a group of individuals depending on the nature of the application. A few of these tasks may be carried out by a particular individual like running, jumping, walking, and sitting through changes in the entire body [1], [2]. Through a particular body part movement, some activities are performed like making hand gestures [3], [4]. Some cases may be done by communicating with objects, such as cooking food in the kitchen [5], [6]. Any abnormal activities like sudden falls [7] are also referred to as HAR. The major successful applications of HAR include ambient assistive living [8], [9], nursing home [10], health monitoring [11], rehabilitation activities [12], surveillance [13], and human-computer interaction [14]. Due to the diverse applications of HAR, it has become a very prominent research issue in the research community. Generally, HAR falls into two categories depending on the types of data such as vision-based HAR and sensor-based HAR [15], [16]. A vision-based technique analyzes the camera data as a video or image format [17], [18] while a sensor-based system interprets the sensors (accelerometer, gyroscope, radar, and magnetometer) data as a time series form [19]. Among the available sensors, the accelerometer is mostly used for HAR due to its low cost, small size, and portability [20], [21].

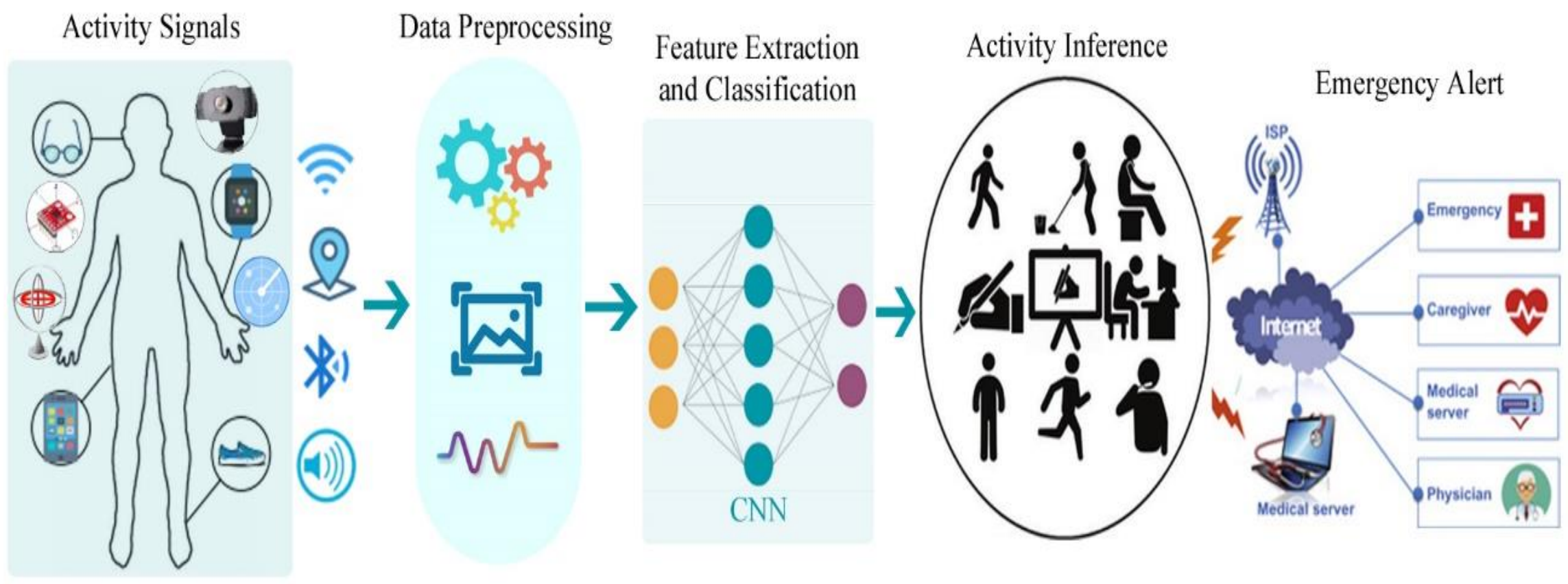

Figure 1. An overall system architecture of CNN-based human activity recognition.

Object sensor like Radio frequency identifier (RFID) tags [22], [23] are utilized in the home environment although it is difficult to deploy. The researches have show n that sensor-based HAR is quite convenient and maintains more privacy compared to vision-based HAR [24], [25]. In addition, vision-based HAR is more influenced by environmental factors like shooting angles, lighting, and overlap between individuals, although it is less expensive to develop [26], [27].

Recently, deep learning algorithms have become more popular due to their automatic feature extraction capability from vision or image data as well as time-series data that enable to learn high-level and meaningful features easily [28], [29]. Deep learning techniques have been generally outperforming traditional machine learning approaches for activity recognition in terms of classification performance measures such as accuracy, precision, recall, and F1 Score [30], [31]. The recognition of human behavior based on deep learning architecture, especially CNN is a composite system that is categorized into several key stages. An overall system architecture with the following stages is illustrated in Figure 1 to realize a CNN-based activity recognition in a short look. The first stage is comprised of the selection and implementation of input devices like sensors, and cameras. Data collection is the next step where an edge device is used to perceive data from input devices and transfers it to the main server through various communication protocols like Wi-Fi, and Bluetooth. The deployment of computing and storage resources at the point where data being collected and processed is referred to as edge computing that incorporates sensors for data perception as well as edge servers for reliable real-time information processing [32], [33]. The feature extraction and selection stage extracts the necessary features from the raw signals; this stage is performed automatically in the case of CNN; no hand-crafted feature extractions are required. This stage contains the CNN architecture or variants of CNN structure for the recognition of activities. The last step includes a notification system through which an agent (human or machine) can be notified.

Although some reviews have been conducted for HAR based on deep learning and machine learning techniques, this article specifically highlights the convolutional neural network-based systems developed for HAR in the scope of multimodal sensors, smartphone, radar data, and vision data. This work provides a comprehensive survey of CNN-based HAR that includes the recent progress as well as an in-depth analysis of the developed systems to serve the research community. The survey discusses the developed systems in terms of performances, strength, weakness, and the used hyperparameters of CNNs. It also provides details on benchmark data sources. Lastly, an analysis is provided to discuss the systems that are best suited for handling all the scenarios of HAR. Additionally, challenges of the existing systems are highlighted with possible suggestions for future research directions.

The rest of the article is arranged as follows. The paper selection process from a large number of existing articles is described in section 2. Section 3 reviews the most recent works that are developed based on CNN architecture. A brief discussion on the available public datasets that are widely used in human activity recognition research is presented in Section 4. The current state of the art with open issues, suggestions for potential future work and applications are discussed in Section 5. Section 6 concludes the paper.

## 2 Selection of papers

In this review work, only the most recent works published in the last nine years in this field are considered. The research was confined to several sources, namely: Google Scholar, PubMed, NCBI, CINAHL, and Web of Science. We used the following keywords for manual web searches and for searching the databases: “CNN based human activity recognition”, “radar based human activity recognition using CNN”, “vision based human activity recognition using CNN, “CNN for multi-modal human activity recognition”, “CNN for smartphone based human activity recognition”, “public datasets for human activity recognition”, “vision datasets for human activity recognition”, “benchmark datasets for human activity recognition, depth image datasets for human activity recognition, and inertial datasets for human activity recognition”. We analyzed the titles and abstracts of the research works found through the search results to eliminate duplicates and to ensure the selected articles fall within the scope of this review work. The specific topics of interest were identified and quantified after thoroughly analyzing each selected article. Figure 2 illustrates the flow of the article selection process for the final review. During the quality assessment process, some articles were removed from the review due to reasons such as: using machine learning algorithms including boosting, random forest, decision tree, and support vector machine (SVM), and using enterprise auto-machine learning solutions.

## 3 Literature Review on Human Activity Recognition

With the rapid growth and performance improvement of deep learning techniques particularly CNN architectures, numerous researchers have adopted them for dealing with HAR problems. This paper is focused on reviewing the four commonly used categories of devices for HAR: i) Human activity recognition from multimodal sensing devices, ii) Human activity recognition from smartphone sensor data, and iii) Human activity recognition from radar signal, and iv) Human activity recognition from image and video signals - as well as the CNN tools that have been used in conjunction with these devices for performing activity recognition. Figure 3 illustrates the human activity recognition systems and their different types based on the sensing devices, signals, or sensing modalities. The multimodal sensing signals based human activity recognition systems generally use data from the following sensing modalities: accelerometer, gyroscope, magnetometer, and barometer. Smartphone-based human activity recognition systems utilize smartphone sensors to collect and classify data. Radar-based human activity recognition systems use various types of radars to collect data. Popular radar variants include Doppler radar, Frequency-Modulated Continuous Wave (FMCW) radar, interferometry radar, and Ultra-wideband (UWB). Vision-based systems collect data through RGB cameras or RGB-Depth (RGB-D) cameras. The recent works for human activity recognition are described below where we illustrated the data collection strategies of each modality and the developed systems for each modality subsequently. In the next few sections, we discuss the four categories of devices and how the CNN based models and tools are utilized to infer human activity from captured data using those devices.

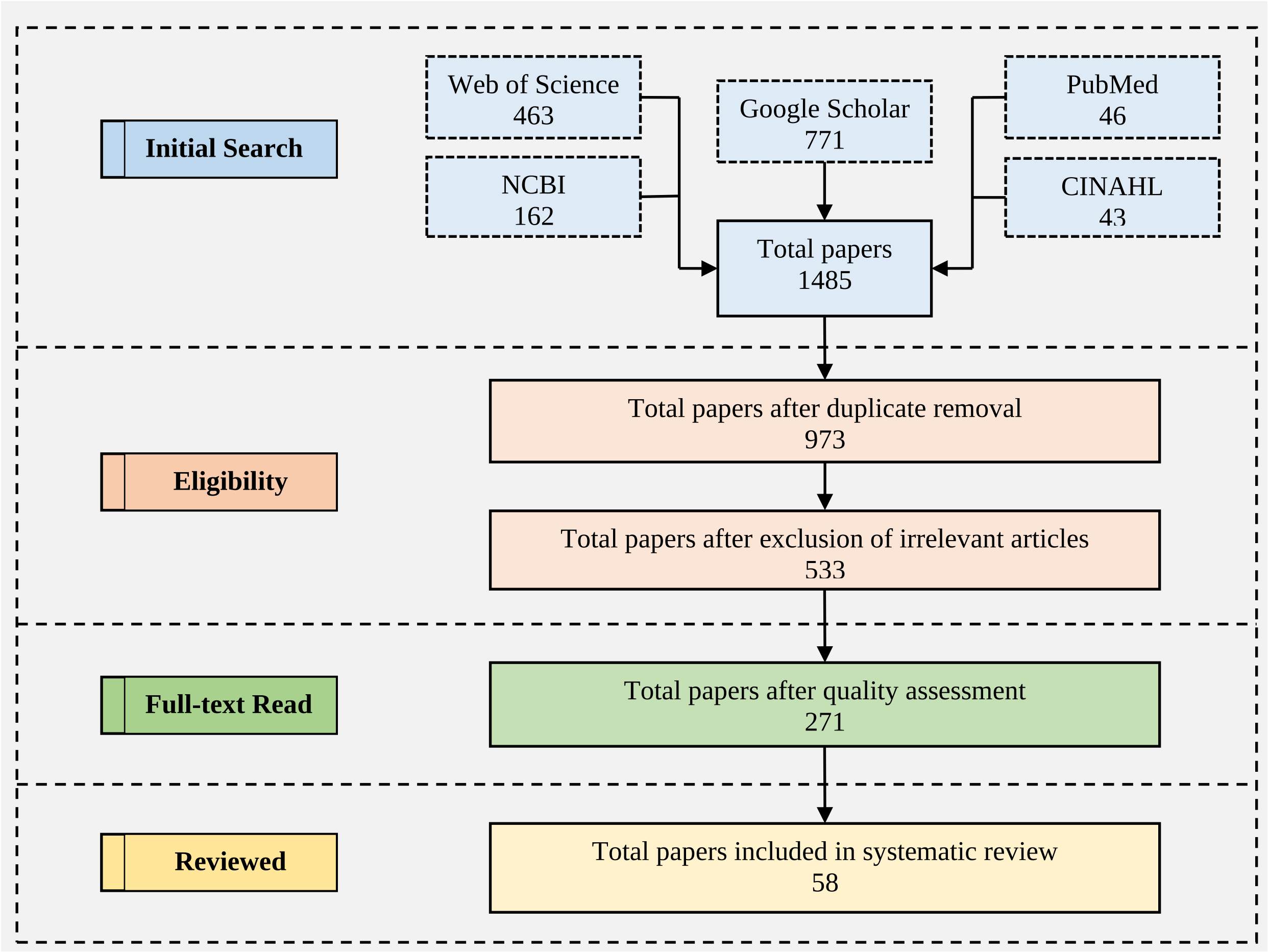

Figure 2. Flow of article selection process for final review.

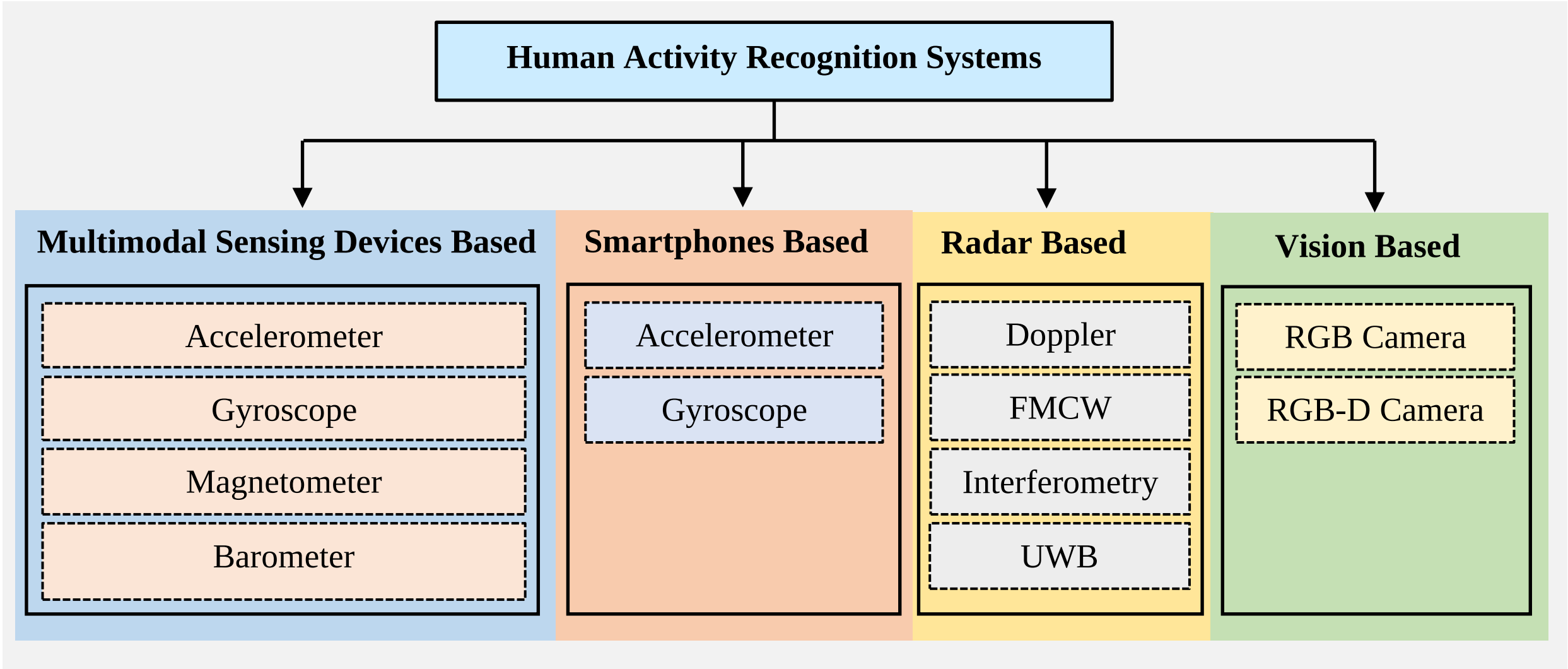

Figure 3. Human activity recognition systems and their modalities.

### 3.1 HAR from Multimodal Sensing Devices

In recent years, the research for activity recognition is focused on the combination of multiple sensor data (accelerometer, and gyroscope) that may increase the performances of the developed system in certain cases [34], [35].

We consider various embedded body-worn solutions such as smart watches, smart necklaces, bands, helmets, and watches containing sensors like 3-D accelerometer, gyroscope, magnetometer, and barometer excluding smartphones as multimodal sensing devices [36]. These solutions are generally smaller than modern smartphones and require less power to run. However, these solutions oftentimes do not have Global Positioning System (GPS) and network connectivity. These devices generally do not require everyday charging. As these devices are always worn or kept on a person by the users, the sensors can record human motion and process them. Human activity generates acceleration and angular velocity. 3-D accelerometer sensor can sense acceleration along the 3-axes. A gyroscope and magnetometer can be used to sense the angular velocity and orientation. A barometer helps in sensing height changes during activities. Combining these properties, multimodal sensing solutions can infer human activities.

In multimodal sensing devices, a big challenge is learning the inter-modality correlations along with the intra-modality data for CNN-based human activity recognition. To solve this issue, some CNN-based approaches have been developed that combine various modalities for the development of single extracted features or ensembles the output of different architectures. The simplest way to handle multimodal sensor data is combining the data from all sensors ignoring the sensor modalities although it has a chance of losing accurate correlations. In order to capture local and spatial dependence over time and sensors respectively, a multi-modal CNN architecture is proposed in [37] that used 2-D kernels in both convolution and pooling layers. As the relationship between the non-adjacent modalities is absent from traditional CNN, the system developed in [38], proposes a novel architecture in which any sequence of signals can be adjacent to every other sequence. Several systems have already been developed that consider each sensing modality initially and then combines them. This architecture provides modality-specific information along with versatile distribution of complexity. An architecture named EmbraceNet was proposed in [39] that processes the sensors' data separately and feeds them into the EmbraceNet. An effective way of fusing multimodal information using this architecture is through docking and embracing structure. A deep multimodal fusion architecture is introduced in [40] that calculates the confidence score of each sensor automatically and combines the features of multi-modal sensors based on the scores.

Some of the frameworks are introduced to minimize the interference between the used sensors by treating each sensor axis independently. In this case, 1-D CNNs are very popular for feature learning of each separate channel. A 1-D CNN architecture, proposed in [41] that omitted the pooling layers to achieve more detailed features where the data from the accelerometer and gyroscope are fed into the network separately. In [42], the acceleration of the accelerometer and gyroscope from seven different body positions are used to produce frequency images that are served as the input of two-stream CNN to learn inter-modality features. Very recently, a few CNN-based systems have been developed to handle univariate time-series data of multichannel where the same sized filter is implemented to all time sequences. In this scenario, the raw signal is converted to a 2-D array by stacking along the modality axis which is then applied to 2-D CNN with 1-D filters [43], [44]. However, the characteristics of all sensor data do not combine externally, but they interact through mutual 1-D filters.

There is another new trend in the area of deep learning called attention mechanism that has become a very popular and frequently used concept in diverse application domains including human activity recognition in recent years [45], [46]. In the current scenarios, the majority of the developed systems used shallow feature learning architectures that could not recognize human activities accurately in real-world situations. To solve this issue, Hamad et al. [47] used a dilated causal convolution with multi-head self-attention for physical activity recognition. During recognition, the multi-headed self-attention is utilized to allow the model to highlight significant and vital time steps rather than irrelevant time steps from the sequential feature space. The proposed architecture obtained an F1 Score of 92.24% from the experimental findings. Wang et al. [48] introduced an activity recognition system that processed weakly labeled information utilizing attention mechanisms. The compatibility between global features and local features is computed using this approach. By weighing their compatibility, the attention mechanism enhanced the salient activity data and suppressed the insignificant and slightly confusing data. The experimental results revealed that the scheme appraised an accuracy of 93.83%. Tan et al. [49] presented faster region based CNN and attention based LSTM for human activity recognition where the CNN extracted the feature as posture vector and the BiLSTM architecture classified the human activities. An attention layer is added between the two BiLSTMs. The developed network obtained precision of 97.02% and recall of 96.83% from the experimental findings.

In the current state of the art, most of the existing frameworks for HAR take the global information of input sequence and avoid local information that demonstrates changes in behaviors, causing the method to be responsive to external factors including occlusion and illumination change. To resolve this problem, some of the studies [50], [51] consider the local spatial features, global spatial features, and temporal features for HAR. Andrade-Ambriz et al. [52] proposed a human activity recognition framework using a temporal convolutional network (TCN) that utilized spatio-temporal features as input of the architecture. The experiment demonstrates that the developed prototype achieved 100% precision and recall for two popular datasets. The scheme shared the activity recognition results to a robot called NAO during real-world environment testing. Zhu et al. [53] introduced a multimodal activity recognition scheme that fused three spatial features such as local, global, and temporal of input signals to classify different human actions. The proposed system divided the input into three segments where the global spatial features are found from the first segment (RGB frame) using a spatial CNN, the local spatial features are extracted from the local blocks utilizing another spatial CNN, and the last segment (optical flow) is utilized to extract temporal features through the use of a temporal CNN. The three architectures are evaluated individually using two benchmark datasets and the final output is obtained using the weighted sum of the three networks. The best accuracy of 94.94% is found from the experimental results for the UCF101 dataset. Gao et al. [54] proposed a framework called DanHAR that combined channel and temporal attention on residual networks to enhance feature representation capability for human activity recognition. The proposed architecture takes a time window as input and sends it to convolutional layers to obtain visual features. This network then generates channel attention through pooling layers (max-pooling and average-pooling) to combine features along the temporal axis. It is found from the experimental results that the proposed architecture obtained an accuracy of 98.85% for the WISDM dataset. In another research, Tang et al. [55] developed a triplet cross-dimension attention model for HAR, which introduced three attention parts to make the cross-interaction between sensor, temporal, and channel dimensions. The performance measure shows that the F1 Score of 98.61% is obtained by the developed system. It is worth mentioning that the system is tested in a real-time environment using a Raspberry Pi prototype.

### 3.2 HAR from Smartphone Sensor Data

Smartphone has become very popular for HAR thanks to the rapid growth of modern technology as it has various built-in sensors for this task [56]. The major problem of the traditional wearable sensing device is that the users should carry an extra device; sometimes they are not willing to carry it, or a few times get forgotten. As almost everyone now has a smartphone, it has become an excellent choice to conduct research using smartphone sensor data, which ensures the portability of the developed systems to a great extent [57].

All modern smartphones contain sensors such as accelerometer, gyroscope, and magnetometer. Smartphones generally require more power than multimodal sensing solutions. However, almost all smartphones require regular daily recharging. Smartphones have more processing powers than multimodal embedded solutions. Smartphones also have GPS and network connectivity, making them viable for transferring data and decisions in various client-server models [58]. Thus, smartphones have access to more sophisticated data-heavy models. Users generally carry smartphones in locations such as thigh, chest, and hand. As human motion generates acceleration and angular velocity, smartphone applications can sense these changes through the embedded smartphone sensors and process them. Smartphones also provide high-level access to sensor data via the Software Development Kit (SDK) of the operating systems. SDK also provides additional support such as always-on applications, notification and alarm systems, and sensor data buffers.

One of the major issues that should be considered is the position of the smartphone as it can be kept in a pant pocket, hands, bags, and shirt pocket. It is evident that due to the various locations of the smartphone, the raw signals change considerably as the movements of the various parts of the body are different. To handle this problem, some of the systems are developed based on the position-independent concept. A smartphone-based position-independent activity recognition system using a CNN was introduced in [59] that used time-domain statistical features. Here, mean-centering is used to convert the raw input to an appropriate form to train the optimum threshold without any bias. A 1-D CNN is introduced in [60] that used smartphone accelerometer data from different body positions like the bag, hand, and pocket for activity recognition. As the data for different body positions are used, the developed system effectively ensures the position-independent property.

It is another big challenge for the smartphone-based system that a deep learning architecture cannot be easily embedded in it as the developed architectures require GPU support that is high cost and power-hungry. To resolve this issue, a few of the systems have been developed that merged the hand-crafted features and deep features for activity recognition. Decreasing filter size is a potential solution to reduce the size of the network that optimizes computing operations. A HAR system is introduced in [61] where the deep features and hand-crafted features are arranged in parallel, and the features are then incorporated into the 1-D CNN architecture. The performance of the developed system has been increased with a small number of computational operations. The system developed in [62] used just one CNN layer and two fully connected layers where the spectrogram features are fed into the network for activity recognition. The experimental findings revealed that the system achieved milliseconds to tens of milliseconds of computing time for a single prediction.

In another study, the authors of [63] proposed an attention-based multi-head architecture for HAR. The developed architecture had three lightweight convolutional heads; each is designed to extract features from collected data using 1-D CNN. The lightweight multi-modal architecture is stimulated with an attention mechanism to improve CNN's representation capability, enabling the automatic selection of significant features while suppressing irrelevant ones. Although two datasets have been utilized here, the highest accuracy, precision, recall, and F1 Score of 98.18%, 97.12%, 97.29%, and 97.20% respectively are achieved from WISDM data. Zhang et al. [64] developed a system that combined the concept of CNN and attention mechanism for activity recognition using the data from a smartphone. Here, the attention is incorporated into multi-head CNNs that facilitate extracting and selecting features efficiently. The proposed scheme achieved accuracy and F1 Score of 96.4% and 95.4% from the experiments. The author of [65] introduced a novel deep learning architecture called LGSTNet, combining the concept of CNN and attention mechanism to recognize human activity from the data of accelerometers and gyroscopes. The activity window is fragmented into various sub-windows in this system, and the local spatial-temporal attributes from those sub-windows are learned using an attention mechanism and CNN. It is evident from the experiments that the obtained F1 Score of the proposed network is 95.69%.

In another research, Nafea et al. [66] presented a HAR framework that used spatial information and temporal information, extracted by CNN with differing kernel sizes and Bi-directional Long Short-Term Memory (BiLSTM), respectively. The retrieved spatio-temporal data were merged in a mixed model that was trained and verified using two benchmark datasets, yielding a 98.53% accuracy, and Cohens Kappa of 98% for the WISDM dataset. Nair et al. [67] developed a temporal convolution network-based network to recognize human activities from raw motion data collected utilizing smartphone sensors. To deal with sequence information with big receptive fields and temporality, dilations and causal convolutions have been developed in this system. The accuracy and F1 Scores of 97.8% and 97.7% respectively are achieved from the experimental results using the encoder-decoder temporal convolutional network.

## 3.3 HAR from Radar Signal

The current research is focused on the device-free approach as it does not include any devices to take while participating in any activities. To ensure a device-free solution, radar-based system shows the best performance due to its insensitivity to daylight and environmental effects as well as contactless-manner [68].

Radar signal-based sensing modality is widely used for stationary surveillance. Radar or radio detection and ranging systems detect both living and inanimate objects through reflection [69]. Radar systems generate intermittent high-frequency radio waves and transmit them in the environment around them. Radio waves are electromagnetic waves that travel at the speed of light. After hitting objects in the environment, radio waves bounce off or reflect from them. Radio systems can receive the reflected radio signal and extract properties of the object including size, shape, distance, and movement from the time required to detect the reflection and the change in frequency due to collision. Radio signal-based sensing modalities deployed in surveillance situations can thus detect stationary objects, humans as well as human motions through the characteristics of the received signal and infer the human activity.

In general, the radar signal is converted to the time-frequency domain which is a separate part from the learning architecture that sometimes does not extract the optimal features. In some cases, the raw signal is transferred to the short-time Fourier transform (STFT) or 2-D matrix and the deep learning architecture treats the 2-D matrix as an optical image. However, the optical image pixels have high spatial correlations, whereas the 2-D radar matrix pixels have a lot of temporal correlations. Hence, treating them as the same is not optimal for classification purposes. To improve the performance of the radar-based systems, some researchers focused on the use of variants of CNN rather than conventional CNN to resolve these issues. An end-to-end network named RadarNet is introduced in [70] where the STFT is substituted by two 1-D convolutional layers. The developed system merged all the steps (micro-Doppler radar data representation, extraction of features, and classification) of HAR in a single architecture. F-ConvNet, another end-to-end architecture is proposed in [71] that used three convolutional layers for multi-scale feature extraction. A novel layer named Fourier layer is proposed here that includes Fourier initializations and two branches of processing for learning the real and imaginary segments individually. In addition, to improve the classification accuracy, dilated convolutions are used. To reduce the computation complexity, an end-to-end network (1-D CNN) is designed in [72] for activity recognition using radar signals. The proposed system used the inception densely block (ID-Block) that is customized for the proposed 1-D CNN where ID-Block is comprised of inception module, network-in-network methods, and dense network.

To solve the issue of limited training data, generative adversarial networks (GANs) are frequently utilized in recent times. In [73], a GAN is developed for HAR using micro-Doppler signatures of radar. While the GAN is trained with the original micro-Doppler images, it generates a lot of similar images like the original that is fed into traditional CNN for training. The use of the increased number of images enhances the performance of the developed system.

### 3.4 HAR from Image and Video data

Due to advances in technology, both RGB and RGB-D cameras are easily accessible and cost-efficient. Vision-based systems are effective for surveillance of large regions in an effective manner.

Image and video sensing modalities are more accessible and easier to setup than radar sensing modalities [74]. Even cheap RGB cameras nowadays have night vision capability through Infrared (IR) sensing. As these systems are stationary solutions, very simple techniques such as background subtraction can be used to localize and monitor motion in the surveilled area. The detected motions can then be passed to CNN models to infer human activity. With the advancement of specialized processors for neural network processing, RGB solutions with built-in neural network processors and network connectivity are available. These systems can capture images or video sequences and process them locally using saved deep learning models in offline mode or can send data to servers running more sophisticated models in online mode.

Although previous research advances were focused on traditional vision-based algorithms, current advances in both deep learning algorithms and hardware enabled us to deploy highly effective deep learning algorithms alongside vision systems. Computer vision-based human activity recognition systems face several challenges such as interclass variation and intraclass similarity, diverse and complex backgrounds, multisubject interactions, group interactions, videos from long distances, and low-quality images. Intraclass variation arises from separate people performing the same action in their own ways. Interclass variation arises from the numerous types of activity we perform in our day to day lives. Vision-based image and video datasets also have various types of backgrounds. Backgrounds differ in lab scenarios as well as in real-world scenarios. Image and video data also suffer from inherent complexities such as pixilation, aliasing, light level differences, viewpoint variations, and occlusions [75]. Video-based human activity recognition datasets are more common than still image-based datasets, as activities are regarded more as a sequence of actions than a one-off scenario.

Most of the available human activity recognition datasets contain video sequences of Activities of Daily Livings (ADLs). The system in [76] introduced an abnormal human activity dataset containing abnormal actions such as coughing, chest pain, faint, vomiting, and taking medication while also implementing a real-time high-speed recognition algorithm based on the You Only Look Once (YOLO) architecture [77]. However, this system only works in cases when a single human is monitored. The system cannot recognize group activities, overlapping objects, and small objects due to the spatial constraints of the YOLO backbone.

While performing human action recognition from video sequences, initial CNN-based research works processed all of the frames of a video for recognizing tasks represented in the video. However, this was inefficient due to the huge time, computation, and memory required to process all the frames. An alternative solution is proposed in [78] where only a selected number of frames of a video are classified instead of all the frames of a video. This drastically reduced the computation time and computation requirements, while also making the system real-time. In general, 30 frames from a video are selected for classification in a deterministic way based on the total number of frames in the video. These selected frames are classified to determine the represented actions and their confidence levels. These confidence levels and actions are averaged to determine the final action and confidence level of the entire video. However, this system works best when a video sequence contains a single action group. When a video contains multiple actions or action groups, determining a single action across an entire video becomes problematic. The CNN methods discussed till now only take the spatial domain characteristics of the videos into consideration. However, temporal domain characteristics can also be extracted from the videos which might act as discriminating features. A two-stream CNN for human action recognition is introduced in [79]. The spatial domain stream is trained on the individual frames of the videos. The temporal domain stream is trained on stacked multiple-frame dense optical flow. The dense optical flow in the temporal domain contains motion data of the objects. The two separate streams are combined by calculating stacked L2 normalized SoftMax scores as features. These features are classified using multi-class linear SVM [80]. The main issue with this method is the huge memory requirements to store all the optical flow data. To reduce the size of the saved data, float point data were converted to integer point data, and the saved data was compressed using JPEG [81]. Despite these measures, the saved optical flow data took huge memory spaces for even smaller datasets. For huge datasets such as the YouTube 1M [82], YouTube 8M [83] this method would require big data storage solutions. An alternate solution for taking the features from the temporal domain into consideration is presented in [84]. As opposed to training two separate 2-D CNN networks in their respective spatial and temporal streams, the system developed in [84] used 3-D convolutions to extract features from both the spatial and the temporal domain. In this way, a single 3-D CNN can be used in place of two-stream two CNN solutions. This also effectively solves the data storage issue of [79]. The performance of this model is comparable to the two-stream two CNN solutions. However, this model requires more labeled data than unsupervised counterparts, and thus, for large video human action datasets, accurate labeling poses a big challenge.

Handcrafted features such as the Histogram of Oriented Gradients (HOG), Histograms of Optical Flow (HOF), Motion Boundary Histograms (MBH) have been historically used for human action recognition [85]. The framework developed in [86] introduces Trajectory-pooled Deep-convolutional Descriptors (TDD), a combination of handcrafted features and deep-extracted features for human activity recognition. The handcrafted features and the features extracted using deep 2-D CNN networks are aggregated based on trajectory constrained pooling. Spatio-temporal normalization and channel normalization are further used to transform the feature maps. The TDD features are highly discriminative and are learned automatically. However, this method performs slightly worse than two-stream 2-D CNN solutions. While the performance of TDD is excellent in the spatial domain, its performance is worse than or comparable to the performance of two-stream solutions in the temporal domain.

## 4 Data Availability

In this section, we explore some of the common benchmark datasets for human activity recognition. The datasets were selected based on the frequency of their usage in the reviewed human activity recognition research works. Most of these datasets are numerical in nature. The datasets contain raw sensor data in some cases and transformed or fusion sensor data in other cases. In almost all cases, the number of test subjects is greater than 10. The data is also collected in different scenarios. In some datasets, the data was collected in laboratory environments. In other cases, the data was collected in out-of-laboratory real-life environments where the data was collected during the real-life events of the subjects. Some datasets are spontaneous, meaning in those cases the subjects were provided with the freedom to perform the activities in their own style. In other datasets, the activities and the steps to perform them were strictly maintained.

Table 1. General characteristics of the datasets

| **Dataset** | **Author** | **Year** | **Number of subjects** | **Scenario** | **Spontaneity** |
|---|---|---|---|---|---|
| mhealth | Banos et. al. [87] | 2015 | 10 | Out-of-lab real-life ADL. | Spontaneous |
| skoda | Zappi et. al. [88] | 2008 | 1 | Activity of assembly line worker in car production scenario. | Not spontaneous |
| ActiveMiles | Ravi et. al. [62] | 2016 | 10 | Out-of-lab real-world ADL. | Spontaneous |
| WISDM v1.1 | Kwapisz et. al. [89] | 2010 | 29 | Out-of-lab real-world ADL. | Not spontaneous |
| WISDM v2.0 | Lockhart et. al. [90] | 2012 | 59 | Out-of-lab real-world ADL. | Not spontaneous |
| Daphnet FoG | Bachlin et. al. [91] | 2010 | 10 | In laboratory ADL data. | Not spontaneous |
| OPPORTUNITY | Roggen et. al. [92] | 2010 | 12 | In laboratory non-ADL data. | Spontaneous |
| UCI | Anguita et. al. [93] | 2013 | 30 | In laboratory ADL data. | Not spontaneous |
| USC-HAD | Zhang et. al. [94] | 2012 | 14 | In laboratory ADL data. | Spontaneous |
| SHO | Shoaib et. al. [95] | 2014 | 10 | In laboratory ADL data. | Not spontaneous |
| YouTube Sports 1M | Karpathy et. al. [82] | 2014 | N/A | Out-of-lab real-life ADL. | Spontaneous |
| YouTube 8M | Abu-El-Haija et. al. [83] | 2016 | N/A | Out-of-lab real-life ADL. | Spontaneous |
| NTU RGB+D | Shahroudy et. al. [96] | 2016 | 40 | In laboratory ADL data. | Not Spontaneous |
| PKU-MMD | Liu et. al. [97] | 2017 | 66 | In laboratory ADL data. | Not Spontaneous |
| UCF101 | Soomro et. al. [98] | 2013 | N/A | Out-of-lab real-life ADL. | Spontaneous |
| KTH | Schüldt et. al. [99] | 2004 | 25 | Combination of both | Not Spontaneous |
| HMDB51 | Kuehne et. al. [100] | 2011 | N/A | Clips from action movies. | Not Spontaneous |
| LIRIS | Wolf et. al. [101] | 2014 | 21 | In laboratory ADL data. | Not Spontaneous |
| UniMiB SHAR | Micucci et al. [102] | 2017 | 30 | In laboratory ADL data | Not Spontaneous |
| PAMAP2 | Reiss et al. [103] | 2012 | 18 | In laboratory ADL data | Not Spontaneous |
| KARD | Gagilo et al. [104] | 2015 | 10 | In laboratory ADL data | Not Spontaneous |
| CAD-60 | Sung et al. [105] | 2012 | 4 | In laboratory ADL data | Not Spontaneous |
| MSR DA | Wang et al. [106] | 2012 | 10 | In laboratory ADL data | Not Spontaneous |
| Ordóñez* SH | Ordóñez et al. [107] | 2013 | 1 | Out-of-lab real-life ADL | Spontaneous |
| Kasteren* SH | Kasteren et al. [108] | 2011 | 3 | Out-of-lab real-life ADL | Spontaneous |

*Note: N/A = Not Available, SH = Smart Home, DA = Daily Activity, Kasteren et. al., and Ordóñez et al. have multiple datasets which were combined for their entries.

Table 2. Sensing modalities and relevant information on the datasets

| **Dataset** | **Number of types of HAR** | **Number of Samples** | **Sensing Points** | **Sensors** | **Position of Sensor** |
|---|---|---|---|---|---|
| mhealth | 12 | 16,740 | 3 | Sh (A, G, M, ECG) | Chest, Right Wrist, Left Ankle. |
| Skoda | 10 | 701,440 | 19 | A | Left Hand, Right hand. |
| ActiveMiles | 7 | 4,390,726 | 1 | S (A, G) | Placement-invariant. |
| WISDM v1.1 | 6 | 1,098,207 | 1 | S (A) | Right Thigh or Left Thigh. |
| WISDM v2.0 | 6 | 2,980,765 | 1 | S (A) | Right Thigh or Left Thigh. |
| Daphnet FoG | 2 | 1,917,887 | 3 | A | Trunk, Thigh, Ankle. |
| OPPORTUNITY | 18 | 27,000 | 72 | A, G, M, Mi, CS | All over the body and environment. |
| UCI | 15 | 10,299 | 1 | S (A, G) | Waist. |
| USC-HAD | 12 | 10,570 | 1 | MN (A, G, M) | Front Right Hip. |
| SHO | 7 | 9,730 | 5 | S (A, G, M, L) | Right Thigh, Left Thigh, Right Waist, Right Upper Arm, Right Wrist. |
| YouTube Sports 1M | 487 | 11,33,158 | 1 | C | Mixed. |
| YouTube 8M | 4,716 | 80,00,000 | 1 | C | Mixed. |
| NTU RGB+D | 60 | 56,880 | 1 | C, D, IR | Mixed. |
| PKU-MMD | 51 | 1,076 | 1 | C, D, IR | Mixed. |
| UCF101 | 101 | 13,320 | 1 | C | Mixed. |
| KTH | 6 | 2,391 | 1 | C | Mixed. |
| HMDB51 | 51 | 7,000 | 1 | C | Mixed. |
| LIRIS | 10 | 828 | 1 | C, D | Mixed. |
| UniMiB SHAR | 9 | 11,771 | 1 | S (A, Mi) | Left Thigh, Right Thigh |
| PAMAP2 | 18 | N/A | 3 | IMU (A, G, M), H | Chest, Wrist, Ankle |
| KARD | 18 | 2160 | 1 | C, D | Mixed |
| CAD60 | 12 | 5440 | 1 | D | Mixed |
| MSR DA | 16 | 23797 | 1 | C, D | Mixed |
| Ordóñez SH | 11 | 50,400 | 12 | IR, F, R | Toilet, Doors, Cupboard, Walls. |
| Kasteren SH | 8 | 74,880 | 23 | IR, F, R, P, T, Me | Toilet, Shower, Doors, Cabinets, Walls, Drawers. |

*Note: Sh = Shimmer wireless sensor, A = 3-D Accelerometer, G = Gyroscope, M = Magnetometer, Mi = Microphone, S = Smartphones, MN = MotionNode, L = Linear Acceleration Sensors, CS = Commercial Sensors, C = RGB Camera, D = RGB Depth Camera, IR = Infrared Sensor, F = Float Sensor, R = Reed Switch, P = Pressure Sensor, T = Temperature Sensor, Me = Mercury Contacts, IMU = Inertial Measurement Unit, H = Heart Rate Monitor. Notation X (Y, Z) indicates the sensing modalities Y and Z are contained within X and X is the main data collection point.

Table 2 provides relevant information related to the datasets such as the number of types of HAR, number of samples, sensing points, the actual sensing modalities, and the position of the sensors. Almost all of the datasets collected sensor data for common activities of daily life such as standing still, jogging, running, jumping, cycling, crouching, climbing stairs, lying down, and walking. The PAMAP2 dataset [103] and the MSR Daily Activity dataset [106] contain some activities that are not generally present in the ADL dataset. Some of the exclusive activities represented in the PAMAP2 dataset [103] are cleaning house, folding laundry, watching TV, and playing soccer. Some of the exclusive activities presented in the MSR dataset [106] are using a vacuum cleaner, cheering up, tossing a paper, and playing guitar. The Kasteren et al. [108] smart home dataset contains activity records during users' using the toilet and shower. The Ordóñez et al. [107] smart home dataset contains activity records during users snacking, sitting idly on a desk, and grooming. The greatest number of sensing points was used in the OPPORTUNITY dataset. Various sensing modalities have been used to collect relevant data. The most commonly used sensing modalities are accelerometer, gyroscope, and magnetometer. The smart home datasets Ordóñez et al. [107] and Kasteren et al. [108] used rare sensors such as reed switches to record usage of doors and cabinets, float sensors to record toilet flushing, and passive IR sensors to record motion. A lot of the datasets used smartphones for data collection and transfer purposes as modern smartphones contain almost all of the commonly used sensing modalities. The sensors were also positioned in various positions of the body. The datasets that used smartphones for data collection generally positioned the smartphone in front thigh positions.

Large-scale vision-based datasets that are collected from the video sharing site YouTube, have a numerous subjects. The system proposed in [82] contains 1 million videos containing various activities, while [83] contains 8 million videos of numerous subjects performing activities across a lot of classes. The position of the camera sensors in these datasets also varies widely. We also consider the datasets containing videos from YouTube as spontaneous in nature, as the videos are made by numerous individuals in various circumstances and as the videos are not staged in most cases.

## 5 Discussions, Open Issues, and Future Works

The research works reviewed in the previous section are the current state of the art in their respective sensing modalities. The detailed discussions and current open issues are described here.

### 5.1 Discussions of reviewed systems

Research works from four sensing modalities were discussed in this review: multimodal sensing devices-based, smartphone-based, vision-based, and radar-based. In a broad sense, both multimodal sensing devices based, and smartphone-based modalities are mostly body-worn or carried by the users. Vision-based and radar modality-based solutions are stationary or surveillance oriented. While smartphone-based solutions provide high-level utility for sensing and processing data alongside network and GPS capabilities, multimodal sensing solutions provide low-power always-on monitoring capabilities without requiring frequent recharging. Smartphone and multimodal solutions often monitor single users at a time, whereas radar and vision-based solutions can monitor multiple persons in an environment. Figure 4 provides a percentile representation of the modalities of the works reviewed in this paper.

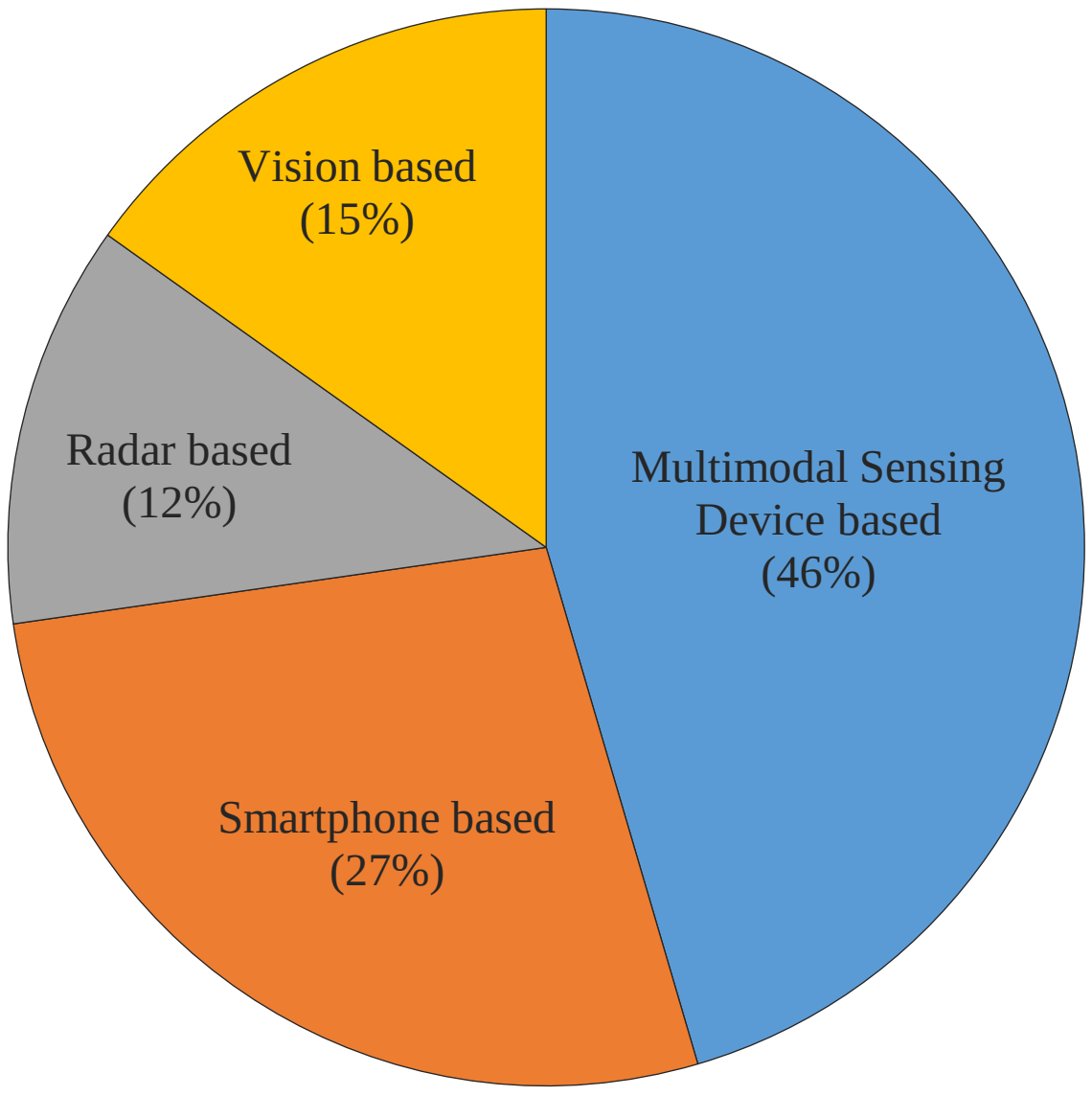

Figure 4. Percentile representation of the modalities in the reviewed studies.

Most of the systems described in this review used benchmark data for their experiments, only a few of the developed systems utilized their collected data by the participants, and almost all the systems used different datasets. Table 3 illustrates the in-depth analysis of the reviewed systems based on the used techniques, performances, strengths, and weaknesses. The highest accuracy (greater than 99%) was found from the reviewed systems [38], [73] for multi-sensing devices data and radar data respectively. In vision-based systems, the highest accuracy was achieved in [76]. However, comparing the systems is difficult due to the differences in the datasets. UCF101 [98] and HMDB51 [100] are the most used datasets in the reviewed vision systems. In general, vision-based human activity recognition works with huge datasets such as the YouTube 1M Sports [82] and YouTube 8M [83] are used. The lowest F1 Score is obtained from [40] for multimodal data. The frequently used dataset is mhealth [87], skoda [88], OPPORTUNITY [92] in this study. A few of the HAR systems [59], [61] provide real-time facilities i.e. ready for practical uses.

Table 3: Depth insight of the reviewed systems highlighting used method, performances, strength, and weakness.

| **Authors** | **Year** | **Dataset** | **Used Method** | **Performance (%)** | **Strength** | **Weakness** |
|---|---|---|---|---|---|---|
| Ha et. al. [37] | 2015 | mhealth [87], skoda [88] | 2-D CNN | Accuracy = 98.29 | The developed framework can capture local and spatial dependency using 2D kernels. | However, the proposed system acquired a slight temporal pattern. |
| Jiang et. al. [38] | 2015 | UCI [93], USC-HAD [94], SHO [95] | DCNN+, DCNN | Accuracy = 99.93 | The framework is able to learn low-level to high-level features. | The computation cost of the developed scheme is comparatively high as the size of the input matrix rises gradually. |
| Choi et. al. [39] | 2019 | Gas sensor arrays dataset [109], OPPORTUNITY [92] | Modified CNN (EmbraceNet) | F1 Score = 91.2 | The use of the docking and embracing layers improve the performance of the developed system. | The system cannot adopt rapidly changing sensors combinations. |
| Choi et. al. [40] | 2018 | Real-time data collected from the participants. | CNN + RNN | F1 Score = 68 | Considered the importance of each modality data based on confidence scores. | The developed scheme somewhat obtained low performance. |
| Yen et. al. [41] | 2020 | UCI open dataset [110], and Recorded data in this work. | 1-D CNN | Accuracy = 95.99, Precision = 96.04, Recall = 95.98, F1 Score = 96.01 | The developed system achieved comparatively good performances for all evaluation metrics. | However, the prototype has not been fully ready yet for application purposes. |
| Lawal et. al. [42] | 2020 | RWHAR [111] | 2-D CNN | Precision = 90, Recall = 90, F1 Score = 90 | Recognized the sensor's position as well as human activities. | The efficiency of the framework is slightly low for practical use. |
| Ha et. al. [43] | 2016 | mhealth [87] | 2-D CNN | Accuracy = 91.9 | Learned modality-specific features through partial and full weight sharing in the lower layers. | Many attempts are being made to achieve minor improvements of the state of the art. |
| Yang et. al. [44] | 2015 | OPPORTUNITY [92], hand gesture dataset [112] | 2-D CNN | Accuracy = 96, F1 Score = 95.5 | An efficient illustration of the local salience of the raw data. | The testing time of the developed system is comparatively high (8 minutes approximately). |
| Hamad et al. [47] | 2021 | UCI [93], Ordonez smart home dataset [107], Kasteren smart home dataset [108] | Dilated causal convolution + multi-head self-attention | F1 Score = 92.24 | The system combines multi-scale contextual data to create insightful feature space. | The operation self-attention mechanism grows quadratically with the input signal, causing training time to increase. |
| Wang et al. [48] | 2019 | UCI [93], Weakly labeled dataset | 2-D CNN + Attention | Accuracy = 93.83 | The system is able to recognize long sequence activity. | The scheme is failed to recognize multiple types of activity. |
| Tan et al. [49] | 2021 | CAD-60 dataset [105] | CNN + attention based LSTM | Precision = 97.02, Recall = 96.83 | The proposed network can extract posture features from skeleton joints from RGB images efficiently. | The system is failed to detect complex activities. |
| Andrade-Ambriz et al. [52] | 2022 | KARD dataset [104], CAD-60 dataset [105], MSR daily activity dataset [106] | 3-D CNN + LSTM | Precision = 100, Recall = 100 | The architecture combines time-motion attributes with the spatial location of human activities. | However, the training and classification time is a little bit high for real-time uses. |
| Zhu et al. [53] | 2019 | UCF101 [98], HMDB51 [100] | Spatial CNN, Temporal CNN | Accuracy=94.94 | The framework can retrieve significant features for recognition across multiple perspectives. | The effects of variation of data in different modes are not demonstrated here. |
| Gao et al. [54] | 2020 | WISDM [89], OPPORTUNITY [92], Weakly labeled dataset | Residual network + Dual attention | Accuracy = 98.85 | The proposed attention mechanism can extract the most crucial data from a long sensor sequence, increasing the comprehensibility of the sensor signal. | The framework combines insignificant sensor modalities, which introduces significant noise. |
| Tang et al. [55] | 2022 | WISDM [89], UNIMIB SHAR [102], PAMAP2 [103], UCI [93], Weakly labeled dataset | ResNet + Attention | F1 Score = 98.61 | The triplet attention obtains a high level of generalization ability. | However, the developed scheme has a large number of parameters that increased the complexity. |
| Almaslukh et. al. [59] | 2018 | RealWorld HAR [113] | 2-D CNN | Precision = 89, Recall = 87, F1 Score = 88 | No generality losses are happened. | Only statistical attributes are considered in lieu of position-independent handcrafted attributes. |
| Lee et. al. [60] | 2017 | $Feature_{10}$, and $Feature_{20}$ dataset | 1-D CNN | Accuracy = 92.71 | Minimized the potential rotational interference of the raw data. | Correlation losses are occurred due to the transformation of vector magnitude from raw signals. |
| Ravi et. al. [61] | 2017 | ActiveMiles [62], WISDM v1.1 [89], WISDM v2.0 [90], Daphnet FoG [91], skoda [88] | 1-D CNN | Accuracy = 98.6 | The inertial sensor information along with complementary data are learned from shallow attributes. | Although the developed system is recommended for real-time application, the latency is slightly high. |
| Ravi et. al. [62] | 2016 | ActiveMiles [62], WISDM v1.1 [89], Daphnet FoG [91], skoda [88] | Temporal CNN | Accuracy = 98.2 | Novel feature generation method that is the sum of the temporal convolutions of the transformed input. | As for resource limitations and the simple strategy of this system, the performance cannot outperform the shallow features based frameworks in some cases. |
| Khan and Ahmad [63] | 2021 | WISDM [89], UCI [93] | 1-D CNN + Attention | Accuracy = 98.18, Precision = 97.12, Recall = 97.29, F1 Score = 97.20 | Squeeze-and-excitation module enhanced the performances of the lightweight architecture. | However, the developed framework cannot recognize complex activities properly. |
| Zhang et al. [64] | 2020 | WISDM [89] | CNN + Attention | Accuracy = 96.4, F1 Score = 95.4 | Manual feature extraction is not required. It learns the necessary features automatically. | The complexity of the developed system is relatively high. |
| Ge Zheng [65] | 2021 | WISDM [89], UCI [93] | 2-D CNN + Attention | F1 Score = 95.69 | The global spatial-temporal features are | However, the features in space domain are not handled here. |
| | | | | | learned efficiently. | |
| Nafea et al. [66] | 2021 | WISDM [89], UCI [93] | CNN + BiLSTM | Accuracy = 98.53, Cohens Kappa = 98 | The relationship between the movement and spatial features is maintained effectively here. | The framework has not experimented with actual users in real-time environment. |
| Nair et al. [67] | 2018 | UCI [93] | Encoder-Decoder TCN, Dilated TCN | Accuracy= 97.8, F1 Score= 97.7 | The raw sensor data is used instead of the more expensive pre-processing. | In complex scenarios, the built-in sensors in smartphones are unable to collect a large number of accurate data. |
| Ye et. al. [70] | 2019 | Real-time data collected for this study. | 1-D CNN | Accuracy = 96.31, Precision = 97, Recall = 97, F1 Score = 97 | The system is able to extract higher-level hidden attributes. | The balance between performances and power requirements is not mentioned. |
| Ye et. al. [71] | 2020 | Data collected using Infineon Sense2GoL Doppler radar. | Fourier CNN | Accuracy = 98.6 | The use of Fourier layer, dilated convolutions, and triplet loss increased the performance. | The micro-Doppler effects somewhat increased the feature extraction complexity. |
| Chen et. al. [72] | 2020 | Real-time data collected from the applicants. | 1-D CNN | Accuracy = 96.1 | The complexity of the developed system is comparatively low. | The developed system somewhat affects with environmental interference. |
| Alnujaim et. al. [73] | 2020 | Data collected from the environment using Doppler radar. | GANs + DCNN | Accuracy = 99.55 | The system showed high performances with synthetically generated data trained with GANs. | The training of multiple GANs would be slightly harder with the increasing number of activities. |
| Gul et. al. [76] | 2020 | Real-time data collected using a RGB camera. | CNN (YOLO) | Accuracy = 96.8, Precision = 90.1, Recall = 88.4, F1 Score = 89.3 | High-speed real-time abnormal human action recognition. | Works only when a single person is performing the task. Does not work well for group activities/small objects/overlapping objects. |
| Shinde et. al. [78] | 2019 | LIRIS [101] | CNN (YOLO) | Accuracy = 88.37, Precision = 89.88, Recall = 88.08, F1 Score = 88.35 | Only classified a small number of selected frames from every video to determine the action in that video, thus | Does not work well in videos that have multiple types of action sequences. |
| | | | | | reducing time, cost, and computation requirements. | |
| Simonyan et. al. [79] | 2014 | UCF101 [98], HMDB51 [100] | CNN + SVM | Accuracy = 88 | Uses a two-stream two 2-D CNN network that is trained on both spatial and temporal domains of video data. Temporal domain takes stacked multi-frame dense optical flow data as input. | Huge storage requirements to store optical flow data even for small datasets despite taking preventive measures such as type conversion and compression. Big data storage and management requires for large datasets. |
| Ji et. al. [84] | 2013 | KTH [99], TRECVID'08 [114] | 3-D CNN | Accuracy = 90.2, Precision = 78.24 | Uses 3-D convolutions instead of two stream 2-D CNN networks solution for taking both spatial and temporal domains of video into consideration. Does not require huge storage solutions. | Required more labeled data than unsupervised approaches which might become a concern in the case of huge datasets. |
| Wang et. al. [86] | 2013 | UCF101 [98], HMDB51 [100] | 2-D CNN | Accuracy = 91.5 (UCF101), Accuracy = 65.9 (HMDB51) | Presents TDD as a robust feature descriptor combining both hand-crafted features and deep-learned features. Performance in spatial domain is equal to or better than other approaches. | Performs slightly worse in temporal domain than other two-stream two 2-D CNN approaches. |

*Note: Some reviewed systems use multiple datasets in their experiments. Here, we considered only the best performance.

Table 4 demonstrates the hyperparameters of the CNN structures that are used to develop each HAR system in this review. Some of the developed systems did not mention the hyperparameters in their work, and some of the frameworks mentioned a few parameters. Among the hyperparameters, the maximum times used batch size, dropout, optimizer, and loss are 64, 0.5, Adam, and cross-entropy respectively.

Table 4: Hyperparameters of the CNN structures used in the reviewed systems.

| Authors | Learning Rate | Number of Epochs | Batch Size | Dropout | Optimizer | Loss |
|---|---|---|---|---|---|---|
| Ha et. al. [37] | N/A | N/A | 60 | 0.5 | N/A | N/A |
| Choi et. al. [39] | 0.001 | 500000 | 64 | 0.5 | Adam | N/A |
| Yen et. al. [41] | 0.00001 | 600 | N/A | 0.5 | Adam | Categorical_crossentropy |
| Lawal et. al. [42] | 0.01 | N/A | N/A | N/A | Adam | N/A |
| Almaslukh et. al. [59] | 0.0003 | N/A | N/A | 0.05 | Adam | N/A |
| Lee et. al. [60] | N/A | 200 | 64 | 0.5 | Adam | Cross entropy |
| Ye et. al. [70] | N/A | 50 | N/A | N/A | RMSprop | N/A |
| Ye et. al. [71] | 0.0001 | N/A | N/A | N/A | SGD | Cross entropy + Triplet loss |
| Chen et. al. [72] | 0.001 | N/A | 64 | N/A | N/A | N/A |
| Alnujaim et. al. [73] | 0.0005 | 3000 | 64 | 0.4 | SGD | Cross entropy |

## 5.2 Open Issues

Although the existing human activity recognition-based techniques are excellent in recognizing atomic and basic activities in single-subject scenarios, they still struggle with HAR in various complex real-life scenarios. Some of the open challenges in HAR systems are presented below.

**Complexity of modeling composite activities:** While basic activities such as walking, running, sitting down are relatively easier to recognize, composite activities such as doing exercises containing multiple routines are significantly harder to model and recognize. There is a significant lack of datasets containing activity data on such concurrent activities.

**Lack of activity data in multi-person scenarios:** The majority of HAR datasets contain data on activities of a single person in the experimental environments. Thus, the majority of the HAR systems are also capable of performing HAR on single-subject scenarios. However, the real world is filled in instances where multi-subject HAR is necessary such as in shops, kitchens, and living rooms, or in the case of multiple subjects involved in atomic activities such as handshaking, and hugging.

**Lack of activity data in group scenarios:** Most of the widely used HAR datasets represent activity data from singular humans performing various activities. Thus, most human activity recognition systems also detect the activity of singular humans. However, humans in real-life perform various activities in groups. There are no datasets that focus on activity data in group activities such as queue in shops, people walking, or jogging together.

**Lack of contextual information in activity data:** As human activity recognition is very closely related to human behavior understanding, context plays a huge role in human activity recognition.

The same activity might be interpreted differently based on the context in which it is performed. Contextual information such as “where” (location) context or “when” (time) context can play huge roles in understanding human activity and behavior. For example, lying down in common resting places such as the sofa or beds can be interpreted as resting action, while lying down in bathrooms or kitchens can be interpreted as fall activity or signs of stroke. Similarly, a person watching tv or walking around the house after midnight might be interpreted as insomniac behavior. Another important contextual information is the repetitiveness of an activity. For example, activities such as eating too many times or too little a day can be interpreted as early symptoms of depression or mental instability. Human-human interaction or human-object interaction is also a great indicator of the meanings behind complex activities. Although contextual information is very important, we have not found datasets or human activity recognition systems that take contextual information into consideration.

**Lack of relevant information in activity datasets:** There are a lot of open datasets related to human activity recognition using various sensing modalities. However, the datasets are not standardized, meaning all of the datasets do not present similar levels of details on the test subjects, test environments, and proper data size. For example, most of the datasets did not provide very important information on the test subjects such as age range, height, and weight.

**Lack of reproducibility of current works:** In a lot of the reviewed works, very specific details on the network architectures and their hyperparameters are not provided. None of the papers provided the code behind the experimentations and model implementations. This makes these systems very hard to reproduce for future experimentation or benchmarking.

**Lack of datasets containing concurrent activity data:** The majority of the HAR systems also perform on the basis that a single human is taking part in a single activity at any given time. However, in real-life scenarios, subjects often perform multi-task and take part in concurrent activities, such as walking while drinking coffee, having snacks, or drinks while reading.

**Lack of benchmark datasets containing data from radar sensing modality:** While there is an abundance of benchmark datasets containing activity data from the various sensing modalities such as the accelerometer, gyroscope, RGB and RGB-D cameras, smartphones, industrial sensors, there is a clear lack of benchmark datasets containing activity data from radar sensing modalities.

**Class imbalance for specific activities:** While data for common tasks such as walking, talking, running, and swimming are very common, data for other abnormal activities such as accidental falls from various positions are very rare even in specialized datasets. Thus, in many cases, a class imbalance exists among the data for different types of activities.

**Inter-activity variability:** Variability is another open challenge for HAR systems. All humans do not perform a single task in the same way. As the CNN models generalize on the training dataset, in test cases, if subjects are performing the same tasks in different ways, then the models become unable to properly recognize them.

**Inherent challenges of underlying technologies***:* The HAR systems based on radar signals have some shortcomings such as the lack of portability, very costly hardware requirements, and environmental interference. Similarly, smartphone-based and wearable sensors-based HAR systems have constraints on wearability. The HAR systems based on multiple sensing modalities are also prone to noise in the data, thus making the data harder to interpret and generalize too. CNN-based systems are also very computationally expensive for both training and inference purposes. While techniques exist to make CNNs more applicable for low-power devices such as embedded systems or edge devices, these methods often affect the overall performance of the systems.

**Difference between real-world data and experimental settings:** Most of the HAR systems were based on data collected in experimental environments. However, real-world activities are very complex and thus harder to model. In addition to composite activities, real-world environments also add complexities such as occlusion, interference, and noise. For example, the type of surface, clothing, previous history of injuries or surgeries affect the activity of individuals in real life.

**Lack of standardization:** Various research works use various testing measures and benchmarks. Some HAR systems use parts of datasets, while others use different testing criteria. Thus, it becomes very hard to perform quantitative comparison among the systems and to perform a proper evaluation.

**Data Privacy or lack thereof:** As HAR systems deal with very confidential real-time data of humans, maintaining the privacy of the collected data is another open challenge. Most of the networked HAR systems are prone to malicious attempts due to the lack of implementation of proper data privacy policies. However, the addition of complex data privacy policies, adding the latencies of the systems, are affecting the performance of real-time HAR systems.

## 5.3 Future Works

HAR systems are continuously evolving along with their underlying technologies. The following research directions can be pursued to further advance the current works in this field.

**Generative models for handling class imbalance:** Recently, generative models such as GANs [115], [116] are being widely used to generate photo-realistic fabricated data. These generative models can be tuned to generate data related to imbalanced classes. This might result in better models that would be capable of recognizing abnormal activities and thus saving lives.

**Future activity prediction:** Future activity prediction is an expansion of HAR that enables the prediction of probable activities by monitored humans. Future activity prediction has applications in law enforcement and driver behavior detection. As human activities are done sequentially in time, using other technologies (ex. Brain-computer interfaces, fMRI, EEG) and mechanisms (ex. attention mechanism) in conjunction with CNNs might result in effective future activity prediction systems.

**Incorporating contextual information with activity data:** Incorporating contextual information alongside sensor data in future open datasets would greatly aid complex human activity recognition systems. For example, incorporating timestamps, location, audio with the relevant sensor data would aid in recognizing complex human actions. Similarly, audio and sentiment analysis systems can be used alongside human activity recognition systems to properly consider environmental contexts.

**Standardization of representation of relevant data in open datasets**: As stated earlier, a lot of the open datasets do not present relevant information such as the age range, height, weight, physical deficiencies, known medical conditions of subjects in the manuscripts. However, different age groups perform similar activities very differently. Similarly, previous medical accidents or surgeries can significantly alter the ability of a subject to perform an action in a specific way. This information can play a huge role in designing criteria-specific or age-specific human activity recognition systems. A standardized data representation system for future human activity recognition datasets should be developed so that the relevant information can be easily accessed and incorporated in designing human activity recognition systems.

**Creating robust human activity recognition systems:** Environmental effects such as the choice of clothing and apparel, type of surface greatly affect the activity of individuals. These changes are also represented in their relevant sensor data. However, the open datasets on human activity recognition are mostly collected in laboratory test environments that do not incorporate these differences. Thus, in the future, research efforts can be focused on developing real-world activity datasets on different environments and robust human activity recognition systems that are impervious to the changes in the environment.

## 5.4 Applications

Human activity recognition systems have widespread applications in various fields [117]–[121]. The main aim of human activity recognitions systems is to understand and analyze classified human actions and to interpret their semantic meaning in different domains. Human activities consist of simple atomic actions [122] such as walking, breathing; complex actions such as dancing, exercising; interpersonal interactions like handshaking, waving; or human-object interaction such as preparing meals, and working in production lines. As human activity recognition has very close ties with human behavior understanding and modeling, human activity recognition systems have application in diverse application domains. In this section, we briefly discuss some of the prevalent application domains of human activity recognition.

### 5.4.1 Healthcare Systems

Human activity recognition systems are widely used in healthcare systems to analyze and interpret patient activities for facilitating healthcare and essential workers to monitor, diagnose, and care for patients [123]. This results in improved accuracy of diagnosis and case, the decreased workload for healthcare staff, increased quality of service received by patients, decreased hospital stays, decreased medical cost, and decreased chances of serious injury. Human activity recognition systems are used in various medical use cases such as automatic fall detection and response systems for detecting accidental falls and providing immediate response services [124]–[127]; respiratory behavior modeling systems to recognize and diagnose sleep disorder [128], cardiovascular diseases, and stroke [129], [130]; medication intake monitoring systems to ensure proper usage of medicine [131], [132]; hand movement monitoring system to recognize and diagnose eating disorders [133], [134]; exercise-aid systems to guide in proper postures during regular exercises; hand gesture recognition system for sign language-based interaction and automatic wheelchair movement [135]–[137].

### 5.4.2 Surveillance Systems

Surveillance systems is another application domain that extensively utilizes human activity recognition systems. Activity recognition systems are used in surveillance scenarios to track and monitor individuals and crowds, thus supporting security personnel to observe and detect suspicious activities and threats. Human activity recognition systems have different use cases in surveillance systems such as gait based long range person recognition and authentication to detect and recognize specific individuals from a long distance based on gait-patterns [138]–[140]; driver drowsiness detection systems to ensure proper driver behavior and to reduce road accidents due to driver inattention [141], [142]; automatic drowning detection systems in swimming pools to save lives of swimmers and to reduce chances of long-term damage [143]; loitering detection systems to detect suspicious loitering behavior or erratic movements of individuals around important public spaces [144]; suspicious activity detection systems to detect violent interpersonal behaviors [145].

### 5.4.3 Entertainment Systems

Human activity recognition systems are widely used in entertainment systems to both monitor and aid referees and players in sports and to interact with computer games in fun ways. Some of the use cases of activity recognition systems in entertainment systems are as follows: accurate automatic timer systems that detect the start and end time of an activity such as swimming [146], diving [147]; pose-estimation systems for detecting and scoring real-life dance moves [148], [149], navigating in 3-D spaces, interacting in virtual environments; movement recognition system for detecting various types of strokes and events during tennis games [150]–[152].

## 6 Conclusion

Human activity recognition systems are essential tools for humanity as they enable the recording of general human activities through different sensing modalities and the monitoring, analysis, and assistance of daily life through capable computing systems. Human activity recognition systems have numerous applications in various important fields such as in healthcare systems, surveillance systems, and entertainment systems. This review work is an exploration of the use of convolutional neural networks in human activity recognition systems through the presented sensing modalities. The major four sensing modalities: multi-sensor-based systems, smartphone-based systems, radar-based systems, and vision-based systems are demonstrated in this review. The different and effective use of various CNN architectures and techniques such as 1-D CNNs, inception blocks, dense blocks, two-stream convolutional networks, 3-D CNNs, specialized features such as trajectory pooled deep convolutional detector are highlighted in this review. The reviewed systems are presented in the light of the issues they solve, their strengths, their weaknesses, and their performance. Available hyperparameters of the reviewed systems are also presented. We also presented brief details on the available public datasets containing data collected through various sensing modalities that are frequently used in the reviewed systems and the field in general. Finally, we discuss the open challenges, the applications as well as some potential solutions.